\documentclass[twocolumn,eqsecnum,preprintnumbers,nofootinbib]{revtex4}
\usepackage{amsmath,amssymb,amsfonts,times,graphicx}

\newcommand{\beq}{\begin{equation}}
\newcommand{\eeq}{\end{equation}}
\def\bea{\begin{eqnarray}}
\def\eea{\end{eqnarray}}

\begin{document}

\title{Entanglement sum rules in exactly solvable models}
\author{Brian Swingle}
\affiliation{Department of Physics, Harvard University, Cambridge MA 02138}

\date{\today}
\begin{abstract}
We compute the entanglement entropy of a wide class of exactly solvable models which may be characterized as describing matter coupled to gauge fields.  Our principle result is an entanglement sum rule which states that entropy of the full system is the sum of the entropies of the two components.  In the context of the exactly solvable models we consider, this result applies to the full entropy, but more generally it is a statement about the additivity of universal terms in the entropy.  We also prove that the Renyi entropy is exactly additive and hence that the entanglement spectrum factorizes.  Our proof simultaneously extends and simplifies previous arguments, with extensions including new models at zero temperature as well as the ability to treat finite temperature crossovers.  We emphasize that while the additivity is an exact statement, each term in the sum may still be difficult to compute.  Our results apply to a wide variety of phases including Fermi liquids, spin liquids, and some non-Fermi liquid metals.
\end{abstract}

\maketitle
\section{Introduction}

A recent exchange of ideas between quantum many-body physics and quantum information science has led to an increased appreciation for the fundamental role of entanglement in quantum matter.  In particular, long range entanglement underlies many of the more interesting states of matter now known experimentally, including Fermi liquids \cite{ee_f1,ee_f2,bgs_f1}, quantum critical points \cite{ee_cft,vidal_crit,eevect,deconf_ee,renyi_free}, and topological phases \cite{topent1,topent2,ent_spec,espec_kitaev,espec_geo}.  In some cases, entanglement is essentially the only completely general probe of such states \cite{topo_kagome1,topo_kagome2} and has led to a clear identification of a topological phase.  Entanglement considerations have also led to a variety of other results, including a new class of variational state \cite{mera,peps,terg} and a classification of phases in one dimension \cite{mps_classify1,mps_classify2}.  The concept of entanglement entropy has played a crucial role in these recent developments, so we first remind the reader about entanglement entropy.

We consider a large quantum system divided into two parts, $A$ and $B$, such that the whole system is in a pure state $|\psi_{AB}\rangle$ e.g. the ground state of a local Hamiltonian.  The entanglement entropy of $A$ is defined as the von Neumann entropy of the reduced density matrix of $A$: $S(A) = - \text{tr}_A (\rho_A \ln{(\rho_A)})$.  Of course this definition makes sense in general, but only when the total system is pure does the entanglement entropy truly measure entanglement between $A$ and $B$.  Typically $A$ and $B$ are spatial regions, but other kinds of entanglement cuts have been considered.  It is also useful to consider the Renyi (entanglement) entropy, defined as
\beq
S_n(A) = \frac{1}{1-n} \ln{\left(\text{tr}(\rho_A^n)\right)},
\eeq
as a generalization of entanglement entropy.  Knowledge of the Renyi entropy for all $n$ is equivalent to knowing the full spectrum of $\rho_R$.

The basic fact about entanglement entropy in local ground states is the area law \cite{arealaw1,arealaw2}.  This law states that the entanglement entropy typically scales like $S(A) \sim L^{d-1}$ where $L$ is the linear size of $A$ and $d$ is the spatial dimension.  However, it must be immediately emphasized that the area law is not completely universal as exceptions known in a variety of gapless systems.  Conversely, although it is not proven, it is believed that the area law holds for all gapped phases of matter (see Ref. \cite{ent_ren_holo} for an renormalization group argument and Ref. \cite{arealaw_log} for a partial result).  The exceptions include conformal field theories in one dimension where $S \sim c \ln{(L)}$ ($c$ is the central charge) \cite{ee_cft,vidal_crit} and Fermi liquids in $d>1$ dimensions where $S \sim L^{d-1} \ln{(L)}$ (as well as other systems with a Fermi surface) \cite{ee_f1,ee_f2,bgs_f1,ee_spinonfs,eethermal_cross}.  Most other known gapped and gapless phases in $d>1$ satisfy the area law.  This central feature of entanglement entropy has already led to a new class of variational states designed to encode the area law and which extend the powerful numerical method known as DMRG \cite{dmrg_orig,dmrg_review}.  In addition, entanglement considerations have led to a fingerprint for topological phases and other strongly correlated systems as well as classification schemes for gapped quantum matter.

However, despite these many advances, entanglement remains fundamentally poorly understood, especially in the context of gapless systems.  Part of this state of affairs stems from the great difficulty, both theoretical and experimental, encountered when trying to compute or measure the entropy.  Fortunately, there are a number of promising new directions that have recently considerably expanded our ability to compute entropies and given us new numerical tests.  Here we focus on what may be called entanglement sum rules.   An entanglement sum rule provides a way to compute the entanglement properties of an interesting phase of matter by dividing it into more elementary components.  A prominent example would be the problem of spin liquids where one often has a description involving matter e.g. spinons coupled to gauge fields.  Both the matter and gauge field physics can be interesting, but only together do they form the unified spin liquid state.  In this context, the prototypical entanglement sum rule is the statement that $S = S_m + S_g$ where $S_m$ and $S_g$ come from matter and gauge fields respectively.  This sum rule is a non-perturbative statement about entanglement entropy that greatly extends our ability to compute entropies and can be very useful for comparing with numerical computations.  Such a rule has previously been obtained in the context of various topological phases \cite{espec_kitaev,deconf_ee} in a certain limit, but our result is much more general.  Of course, not every spin liquid state admits an entanglement sum rule, but many interesting states do.  We also emphasize that the notion of an entanglement sum rule is general and is not restricted to spin liquids.

In this work we prove a general entanglement sum rule for a wide class of exactly solvable models that include in various limits free fermions, the toric code, and free fermions coupled to the toric code.  A general model encompassing all these limits was recently introduced in Ref. \cite{ometal} to describe so-called orthogonal metals.  In the context of the exactly solvable model our sum rule applies to the full entanglement entropy, but more generally, our sum rule indicates that the universal terms in a variety of phases and phase transitions will be additive in the sense described above.  Note that what terms are considered universal depend on the nature of the phase.  An important advance over the results in Refs. \cite{espec_kitaev,deconf_ee} is that we can also vary the gauge field dynamics (both previous computations were down in an extreme deconfined limit where the gauge field does not fluctuate).  Below we describe an exactly solvable model, introduced in Ref. \cite{ometal} in the context of non-Fermi liquid metals, and prove the entanglement sum rule for this model.  We also generalize the model while preserving the sum rule.  Ultimately our sum rule applies to a huge variety of physical states including Fermi liquids, non-Fermi liquid metals, spin liquids, deconfined critical points, and much more.

\section{Exactly solvable model}

Consider a square lattice with fermions (electrons) $c_r$ on sites and Ising spins $\sigma^z_{rr'}$ on links.  Following Ref. \cite{ometal} the Hamiltonian is taken to be
\bea \label{orthometal}
&& H = - w \sum_{<rr'>} c_r^\dagger \sigma^z_{rr'} c_{r'} - \mu \sum_r c_r^\dagger c_r \cr \nonumber \\
&& - g J \sum_{<rr'>} \sigma^z_{rr'} - J \sum_r (-1)^{c_r^\dagger c_r} \prod_{<r'r>} \sigma^x_{rr'} \cr \nonumber \\
&& - U \sum_p \prod_{<rr'> \in p} \sigma^z_{rr'}.
\eea
The $J$ term is a product over all links sharing a vertex $r$ while the $U$ term is a product over all links in given plaquette $p$.  We will see that this Hamiltonian describes a (highly fine tuned) transition from a Fermi liquid to an orthogonal metal as a function of $g$.  Indeed, this Hamiltonian is actually exactly solvable for all values of the parameters, but first let us get a sense of the physics as a function of the coupling $g$.

When $g \gg 1$ the spins want to polarize as $\sigma^z_{rr'} = 1$.  The fermions then decouple from the spins and are simply described by a free Fermi gas with hopping $w$ and chemical potential $\mu$.  In the opposite limit, when $g \ll 1$, the spin Hamiltonian is that of the toric code or of $Z_2$ gauge theory with gapped matter, and the fermions $c_r$ couple minimally to the gauge field.  Thus in this limit we have again a free Fermi surface coupled to the tensionless limit of a $Z_2$ topological phase.  The topological structure tells us that there has to be transition between the $g \gg 1$ and $g \ll 1$ phases.

The $U$ term commutes with everything else in the Hamiltonian and hence we may consider only the subspace in which $\Phi_p = \prod_{<rr'>\in p} \sigma^z_{rr'} = 1$ for all $p$.  In our interpretation above, $\Phi_p$ is the $Z_2$ flux through the plaquette $p$.  Within this subspace the $\sigma$ variables are constrained, so we can introduce new variables via
\beq
\sigma^z_{rr'} = \tau^x_r \tau^x_{r'}
\eeq
and
\beq
\prod_{<r'r>} \sigma^x_{rr'} = \tau^z_{r}.
\eeq
The commutation relations of these operators are preserved by this identification provided the $\tau$ variables also obey the standard Pauli commutation relations.  Physically, $\tau^z_r$ measures the $Z_2$ charge of a site while $\tau^x_r \tau^x_{r'}$ creates $Z_2$ charges at $r$ and $r'$.  Note that ordinarily we would need to stretch a Wilson line between two such charges, but because the flux $\Phi_p$ is exactly one, no Wilson line is necessary.

If we now define
\beq
f_r = \tau^x_r c_r,
\eeq
\beq
\tilde{\tau}^x_r = \tau^x_r,
\eeq
and
\beq
\tilde{\tau}^z_r = (-1)^{c_r^\dagger c_r} \tau^z_r = (-1)^{f_r^\dagger f_r} \tau^z_r ,
\eeq
then the $\tilde{\tau}$ and $f$ variables all commute and the Hamiltonian is
\bea
&& H = -w \sum_{<rr'>} f_r^\dagger f_{r'} - \mu \sum_r f_r^\dagger f_r \cr \nonumber \\
&& - g J \sum_{<rr'>} \tilde{\tau}^x_r \tilde{\tau}^x_{r'} - J \sum_r \tilde{\tau}^z_r.
\eea
Thus we have decoupled fermion ($f$) and tranverse field Ising ($\tilde{\tau}$) systems and hence the full Hamiltonian is solved in terms of these two models.  We have also dropped the $U$ term since it is simply an additive constant within the constrained Hilbert space.

We see immediately that $ff$ correlations are identical to that of a free fermion model for all $g$.  In particular, since $n_r = c_r^\dagger c_r = f_r^\dagger f_r$ it follows that the physical density-density correlator is given by the free fermion result for all $g$.  Thus the number fluctuations in a region $R$, defined as the variance of the operator $N_R = \sum_{r\in R} n_r$, is given by the free fermion value for all $g$.  Furthermore, the thermodynamics of the model exactly factorizes so that
\beq
S(T) = S_f(T) + S_{Ising}(T)
\eeq
where $S(T)$ is the thermal entropy.  As a technical subtlety, this entropy result only applies if we first send $U\rightarrow \infty$ so that no states with $\Phi_p \neq 1$ are excited since otherwise the $\tau$ variables are insufficient to describe the full Hilbert space.  Indeed, since the Ising part at low temperatures is either gapped or at worst a 2d CFT with $S_{Ising} \sim T^2$ it immediately follows that the thermal entropy is dominated by the $f$ Fermi surface at low temperatures.

\section{Entanglement entropy}

We now turn to the entanglement entropy.  How we compute the entropy depends on what cut we choose to make.  For example, if we regard $c$ and $\sigma$ as the local degrees of freedom, then we cannot make a cut in the $\tau$ variables since these are non-locally related to the $\sigma$ variables.  To begin, let us determine the ground state $|0\rangle$ of the system.  Clearly we have
\beq
|0\rangle = | f(w,\mu) \rangle |\tilde{\tau}(g)\rangle
\eeq
where $|f\rangle $ and $|\tilde{\tau}\rangle$ are the ground states of the decoupled sectors.  To compute the entropy due to a $c,\sigma$ cut we must rewrite this state in terms of the appropriate variables.

As a first step, let us consider the relation between $\tau$ and $\tilde{\tau}$.  Since $\tilde{\tau}^z = (-1)^{c^\dagger c}\tau^z$, we can consider the unitary
\beq
U = (1-c^\dagger c) \otimes 1 + c^\dagger c \otimes \tau^x
\eeq
which satisfies
\beq
U^\dagger = U,
\eeq
\beq
U^2 = 1,
\eeq
and
\bea
&& U \tau^z U^\dagger = U \tau^z U \cr \nonumber \\
&& = (1-c^\dagger c) \otimes \tau^z + c^\dagger c \otimes (-\tau^z) \cr \nonumber
&& = (-1)^{c^\dagger c}\tau^z.
\eea
Using $\bigotimes_r U_r$ we can convert all $\tau$ variables into $\tilde{\tau}$ variables, and hence it follows that
\beq
|0 \rangle = \left(\bigotimes_r U_r\right) |c(w,\mu)\rangle |\tau(g)\rangle,
\eeq
or in other words, the state in the $c,\tau$ variables differs from that of the $f,\tilde{\tau}$ variables by a local unitary transformation (product of CNOT gates).  Indeed, we see that if we made a cut in the $c,\tau$ variables then the entanglement entropy would obey $S = S_c + S_\tau$ since $\bigotimes_r U_r$ is explicitly a product of single site unitary operators.

The next step is to relate the $\tau$ variables to the $\sigma$ variables.  The state $|\tau(g)\rangle$ corresponds to some particular state $|\sigma \rangle$ as a function of the $\sigma$ variables, but we must also account for the unitary $V = \bigotimes_r U_r$.  If we require the number of fermions to be even, then every term in the action of $V$ on $| c \rangle | \sigma \rangle $ amounts to a product of $N_c$ $\tau^x$ or $N_c/2$ pairs of $\tau^x$ operators ($N_c$ is the total number of fermions).  Furthermore, every pair of $\tau^x$s can be written as a string of $\sigma^z$ operators and hence every term in the action of $V$ on the state is the product of a large number of $\sigma^z$ operators.  Note also that the choice of this product is largely arbitrary since the ground state obeys $\Phi_p =1$ for all $p$.  The density matrix is therefore
\beq
\rho_R = \text{tr}_{c,\sigma,\bar{R}}(|0\rangle \langle 0|) = \text{tr}_{\bar{R}}(V |c\rangle |\sigma \rangle \langle c | \langle \sigma | V^\dagger ).
\eeq

Now consider an arbitrary term in the expansion of $V$ of the form
\beq
\prod_{r\in C_R} n_r \tau^x_r \prod_{r\in R-C_R} (1-n_r) \prod_{r\in C_{\bar{R}}} n_r \tau^x_r \prod_{r\in \bar{R}-C_{\bar{R}}} (1-n_r)
\eeq
where $C_R$ and $C_{\bar{R}}$ denote collections of sites in $R$ and $\bar{R}$.  As above, we can replace products of $\tau^x$ with products of $\sigma^z$, and in the case when $|C_R|$ is even, we can choose the $\sigma^z$ configuration to lie entirely within $R$ and $\bar{R}$.  Similarly, in the case when $|C_R|$ is odd we must have one $\sigma^z$ on the boundary connecting $R$ and $\bar{R}$.  Write $|c\rangle =|c,0\rangle + |c,1\rangle$ where $0,1 = N_R \mod{2}$ labels the parity of $N_R$ (which is also the parity of $N_{\bar{R}}$).  Within a given parity subsector, we can completely factorize the action of $V$ (by choosing a suitable pattern of $\sigma^z$s) into unitaries acting on $R$ and $\bar{R}$ separately,
\beq
V |c,x\rangle = V_{R x} V_{\bar{R}x} |c,x\rangle.
\eeq
This fact enables us to write the state as
\beq
V_{R0}V_{\bar{R}0} |c,0\rangle |\sigma\rangle + V_{R1}V_{\bar{R}1} |c,1\rangle |\sigma\rangle
\eeq
where $V_x$ preserves the parity of $|c,x\rangle$.

Let us now compare the reduced density matrix of $R$ with and without the action of $V$.  Without $V$ we have
\beq
\text{tr}_{\bar{R}}\left((|c,0\rangle + |c,1\rangle )(\langle c,0| + \langle c,1|)\otimes |\sigma \rangle \langle \sigma |\right)
\eeq
which reduces to
\beq
\text{tr}_{\bar{R}}\left((|0\rangle \langle 0|+|1\rangle \langle 1|)\otimes |\sigma \rangle \langle \sigma |\right) \equiv \rho^{(0)}_{R0} + \rho^{(0)}_{R1}
\eeq
since the cross terms cancel.  Now with $V$ we have
\bea \label{rhoRV}
&& \text{tr}_{\bar{R}}\left(V_{R0}|0\rangle \langle 0|\otimes |\sigma \rangle \langle \sigma |V_{R0}^\dagger +V_{R1}|1\rangle \langle 1| \otimes |\sigma \rangle \langle \sigma | V_{R1}^\dagger\right) \cr \nonumber \\
&& = V_{R0} \rho^{(0)}_{R0} V^\dagger_{R0} + V_{R1} \rho^{(0)}_{R1} V^\dagger_{R1}.
\eea
Note that once again the cross terms cancel since, for example, we can perform the trace in the number basis with no matrix elements between different parity sectors.  However, we now see that since the state of $R$ before $V$ acted was diagonal in parity, it follows that the state after $V$ acted can be related to the state before $V$ acted by a unitary transformation.  Define $V_R$ to be
\beq
\left(
  \begin{array}{cc}
    V_{R0} & 0 \\
    0 & V_{R1} \\
  \end{array}
\right)
\eeq
where the two components refer to the parity sector.  Computing
\beq
V_R (\rho^{(0)}_{R0} + \rho^{(0)}_{R1} ) V_R^\dagger
\eeq
we obtain
\beq
V_{R0} \rho^{(0)}_{R0} V^\dagger_{R0} + V_{R1} \rho^{(0)}_{R1} V^\dagger_{R1}.
\eeq
since $\rho^{(0)}_{Rx}$ has a definite parity of $x$.  To summarize, we have shown that the action of $V$ on the state $|c \rangle |\sigma\rangle $ reduces, after the trace over $\bar{R}$, to a unitary transformation of the density matrix of $R$.  Hence all spectral data is unmodified by the application of $V$ despite the non-local transformation between $\sigma$ and $\tau$.

Our final result is then $S_n(R) = S_n(R;c)+S_n(R;\sigma)$ since the state $|c\rangle |\tau \rangle$ is factorized.  Note, however, that the trace is still over the $\sigma$ variables (not the $\tau$ variables) and hence we must write the $|\tau \rangle$ state in terms of $\sigma$ variables.  This transformation is critical since in the $\tau$ variables the state is a product at both large and small $g$, but this is not true for the physical $\sigma$ variables.  In words, we find that the total Renyi entropy is simply the sum of the Renyi entropy of fermions with bandstructure specified by $w,\mu$ (with no coupling to $\sigma$) and the Renyi entropy of the $\sigma$ model (with no coupling to $c$).  We emphasize again that while this means the fermionic component of the entropy may be computed directly from the state $|f \rangle$, we must still convert from the non-local $\tau$ variables to the local $\sigma$ variables to compute the $\sigma$ contribution to the entropy.

The Ising part of the entanglement entropy never scales faster than $L$ ($L$ is the linear size of $R$), and thus the entanglement entropy has an $L \ln{(L)}$ term determined by the $f$ Fermi surface for all $g$.  The entropy of the system at $g=\infty$ is simply given by the $f$ Fermi surface while the entropy at $g=0$ is given by $S = S_f + (|\partial R|-1)\ln{2}$ in accord with the results of Ref. \cite{espec_kitaev,deconf_ee} (as usual, we do the splitting over $\sigma$ variables as in Ref. \cite{topent2}).  Our argument works at the level of a unitary transformation and since the parity structure we used in our proof is also a property of excited states, we see that the full entanglement-thermal crossover function $S_n(R,T)$ is also exactly additive \cite{eethermal_cross}.  However, we emphasize again that we must first send $U\rightarrow \infty$ for the thermal results to hold in this simple form.

Within the Fermi liquid phase we have thus partially confirmed the universality claimed in Refs. \cite{bgs_f1,bgs_f2} since we have exhibited a Fermi liquid state with with variable quasi-particle residue in which the entropy is exactly the free fermion result as regards the $L \ln{(L)}$ term.  Furthermore, because the $\sigma$ variables are gapped within the Fermi liquid phase, all of the mutual information calculations of Ref. \cite{bgs_f3} also go through in the Fermi liquid state (the generalization of our results above to multiple regions is trivial if the $\sigma$ variables are gapped).  Recently, an interesting numerical calculation \cite{numerical_fs_ee} of the Renyi entropy in Fermi liquids has also largely confirmed the predictions of \cite{bgs_f1,bgs_f2,bosonization_fs_ee} concerning the universality of the Widom formula with small deviations observed only at very strong interactions.  The precise origin of these discrepancies is not yet understood.

\section{Generalized solvable models}
We may modify the Hamiltonian in Eq. \ref{orthometal} in many ways to produce other exactly solvable models.  The simplest modification is to replace the simple nearest neighbor hopping model for the $c$ fermions with a more complex bandstructure including terms like $c^\dagger_r c_{r'}$ with $r$ and $r'$ two or more links apart.  To keep the model exactly solvable each such term should be augmented with a factor of $\prod_{<rr'>\in \gamma} \sigma^z_{rr'} $ where $\gamma$ is any link path that begins at $r$ and ends at $r'$.  We can easily check that because $\Phi_p=1$ in the ground state the choice of path $\gamma$ is irrelevant, and using the change of variables
\beq
\sigma^z_{rr'} = \tau^x_r \tau^x_{r'}
\eeq
we see that
\beq
\prod_{\ell \in \gamma} \sigma^z_{\ell} = \tau^x_r \tau^x_{r'}
\eeq
and hence each $c$ fermion can still be naturally combined with a $\tau^x$ to produce an $f$ fermion.  Thus we can produce any shape Fermi surface we want as well as Dirac cones and other structures.

As an example of the entanglement sum rule in such a model, consider the case where the fermions form Dirac cones.  In this case, the entropy of the fermion component is expected to scale as $S \sim a L - b$ with $b$ universal (see Ref. \cite{renyi_free} for a calculation of $b$ for a disk).  Similarly, for $g\ll 1$ the entropy of the gauge field component is $S \sim a' L - b'$ with $b' = \ln{(2)}$.  Our results imply a universal term for the coupled system of $b_{\text{Dirac}} + \ln{(2)}$ which should be compared with the results in Refs. \cite{espec_kitaev,deconf_ee}.

We can also include density-density interactions for the $c$ fermions since we have already seen that $c^\dagger c = f^\dagger f$.  The $f$ fermion model may not be so easily solved, but the decomposition of the entanglement and thermal entropies still hold.  Furthermore, we can substitute bosons, call them $b$, for the $c$ fermions without changing the story, and we can of course include boson-boson interactions as well.  It is even possible to add pairing terms for the fermions, again with the appropriate factors of $\sigma^z$.  There is one subtlety, however, since if a single boson condenses then the $Z_2$ gauge structure is immediately lost.  In this case, our additivity result no longer applies.  Of course, a fermion (or boson) pair condensate is perfectly consistent with the $Z_2$ gauge structure since neither carry $Z_2$ charge.

As an example of the failure of additivity, suppose the boson wavefunction before the application of $V$ (or really its analog in the bosonic case) is simply $|b \rangle = \bigotimes_r |\beta \rangle_r$ where $b|\beta \rangle = \beta | \beta \rangle$.  Suppose also that the gauge field sector is in the $g\gg 1$ limit with the state $|\sigma \rangle$ given by a sum over all closed loops in the $\sigma^x$ basis.  The transformation which attaches the ends of strings to the bosons (the analog of $V$) is
\beq
V_b = \bigotimes_r \left( \frac{1+(-1)^{b^\dagger_r b_r}}{2} + \frac{1-(-1)^{b^\dagger_r b_r}}{2} \tau^x_r \right).
\eeq
Following the discussion above, if $b$ were not condensed then we would divide the wavefunction into even and odd parity sectors and the argument would proceed.  However, since parity is no longer well defined due to the condensate, it follows that the state of $R$ before and after $V_b$ acted differ in entropy.  This is because, as shown in Eq. \ref{rhoRV}, the application of $V$ partially decoheres the state of $R$ into parity sectors thus adding entropy.  Indeed, assuming that the boson number fluctuates a lot, each parity sector will have roughly the same weight and $V_b$ will add $\ln{2}$ entropy (associated with the parity bit), exactly canceling the topological contribution of $-\ln{2}$ and indicating that the state is topologically trivial.

We can also trivially generalize the model to higher dimensions.  Returning the Eq. \ref{orthometal}, nearly every term generalizes immediately to higher dimensions.  Let us take a cubic lattice in $d=3$ as an example.  The fermion hopping terms obviously generalize as does the $gJ$ term.  The appropriate generalization of the $J$ term is to associate with each site $r$ the operator
\beq
\prod_{<r'r>} \sigma^z_{rr'}
\eeq
where the product is over the six links emanating from $r$.  Finally, the $U$ term is unchanged in form, but we must now sum over all faces of the cubic lattice. With these simple modifications the algebraic structure of Eq. \ref{orthometal} is preserved and all subsequent changes of variable go through as in $d=2$.

Another direction for generalization is to consider a $Z_n$ gauge field.  We introduce a link variable $Z$ satisfying $Z^n=1$ and a conjugate variable $E$ which satisfies $e^{2\pi i E/n} Z = e^{2\pi i/n} Z e^{2\pi i E/n}$.  $Z$ may be be understood as incrementing $E$ by one, while $E$ may be viewed as measuring the electric flux along a link.  Now consider a square lattice in $d=2$ with Hamiltonian
\bea
&& H = -w \sum_{<rr'>} c_r^\dagger Z_{rr'} c_{r'} - \mu \sum_r c_r^\dagger c_r \cr \nonumber \\
&& - g J \sum_{<rr'>} (Z_{rr'}+Z_{rr'}^*) - J \sum_r e^{2 \pi i c^\dagger_r c_r /n}\prod_{<r'r>} e^{2\pi i E_{rr'}/n} \cr \nonumber \\
&& - U \sum_p \left(\prod_{<rr'>\in p} Z_{rr'} + \prod_{<rr'>\in p} Z^*_{rr'}\right) .
\eea
At large $g$ we have $Z_n$ gauge theory in the tensionless limit or the equivalent string net model while at small $g$ the model favors $Z_{rr'} = 1$ on all links.  Repeating our analysis above with parity replaced by the fermion number $\mod n$ gives an essentially identical entanglement sum rule.  There is, however, a greater constraint on the allowed fermion states, since we must leave the $Z_n$ gauge structure unbroken.  The sum rule will break down for fermion pair condensates but will remain valid provided we condense a multiple of $n$ fermions (or bosons).

\section{Discussion}
In this work we have derived an entropy sum rule for a wide class of exactly solvable models.  Furthermore, it is clear that this technique has not been exhausted and may give further useful insights into the entanglement structure of quantum matter.  There are many simply extensions of our arguments here, some of which we have already sketched, but it is probably also true that other classes of models admit structures like the entanglement sum rule.

For the exactly solvable models we considered, we have proven exact additivity for the full entropy.  More generally, however, our results imply that the universal terms will add in any phase smoothly connected to one describable by our exactly solvable models.  These phases include the Fermi liquid, the orthogonal metal \cite{ometal}, various kinds of spin liquids, exotic superconductors, orthogonal nodal metals, and much more.  Moreover, we have the added ability to follow the entropy through various quantum critical points, and provided the critical point is not heavily fine tuned (as unfortunately is the case with the orthogonal metal), universal terms at the critical point should add as well.

We also give an example of a broader application of our ideas.  It was noted in Ref. \cite{ometal} that Luttinger's theorem, when take to refer to singularities in the electron spectral function, is not satisfied in the orthogonal metal e.g. the electron spectral function sees an overabundance of critical surfaces relative to the $f$ Fermi surface.  Of course, Luttinger's theorem only applies in its original form to Fermi liquids, so nothing is surprising about the observation of Ref. \cite{ometal}, but it has been possible to preserve the counting embodied in Luttinger's theorem in other phases.  Here we see that the Luttinger count may be upheld provided we use the entanglement entropy to measure the hidden $f$ Fermi surface.  This is intriguing in that it makes contact with recent attempts to measure hidden Fermi surfaces in compressible states using entanglement entropy in a holographic context \cite{PhysRevB.85.035121}.  What we have shown is that the entropy does exactly capture the hidden Fermi surface in the compressible orthogonal metal.

Finally, let us conclude by noting that many recent experiment measurements on the organics \cite{Yamashita04062010,PhysRevB.77.104413} and other two dimensional magnetic systems have revealed evidence of a spin liquid ground state, possibly one including a Fermi surface of spinons.  However, the identification of the precise spin liquid state has remained elusive with various experiments pointing in different directions.  Thus more detailed information is needed, and it may be that the detailed information on entanglement presented here when combined with numerical work can help elucidate the precise nature of these exciting new phases.

\textit{Acknowledgements.}
BGS is supported by a Simons Fellowship through Harvard University and acknowledges helpful conversations on some of these topics with S. Sachdev, L. Huijse, and T. Senthil.

\bibliography{ee_orthometal}
\end{document}